# Intrinsic defects as a source of *n*-type conductivity in CrSBr

Timur Biktagirov[1*], Wolf Gero Schmidt[1], Karl Jakob Schiller[2], Michele Capra[2], Jonah Elias Nitschke[2], Lasse Sternemann[2], Mira Sophie Arndt[2], Giovanni Zamborlini[2,3], Anna Isaeva[2], Mirko Cinchetti[2*]

[1] Department of Physics, Paderborn University, Warburger Str. 100, 33098 Paderborn, Germany

[2] Department of Physics, TU Dortmund, Otto-Hahn-Str. 4, 44227 Dortmund, Germany

[3] Institute of Physics, Karl-Franzens-Universität Graz, Universitätsplatz 5, 8010 Graz, Austria

*Corresponding authors:
Timur Biktagirov: timur.biktagirov@uni-paderborn.de
Mirko Cinchetti: mirko.cinchetti@tu-dortmund.de

**ABSTRACT.** Understanding and controlling native defects is essential for unlocking the full potential of two-dimensional magnetic semiconductors. Here, angle-resolved photoemission spectroscopy (ARPES) and first-principles calculations are used to explore the electronic properties of bulk CrSBr. ARPES measurements reveal clear signatures of conduction band filling in as-grown crystals, indicative of unintentional doping. An analysis of intrinsic defects based on density functional theory (DFT) identifies chromium interstitials ($Cr_i$) stabilized between CrSBr layers as the most favorable shallow donors. Bromine-on-sulfur antisites ($Br_S$) and bromine vacancies ($V_{Br}$) are also found to act as potential donors, albeit with deeper ionization energies. Our results shed light on the origin of unintentional *n*-type doping of CrSBr and pave the way for new strategies for defect control and electronic property tuning in this van der Waals magnet.

# Introduction

Two-dimensional (2D) magnetic semiconductors are promising for spintronic and quantum information technologies due to their unique electronic, magnetic, and optical properties. Among these materials, chromium sulfur bromide (CrSBr)[1] stands out for its bulk's robust air stability, high magnetic ordering temperature ($T_N \approx 132$ K)[2,3], and sizable direct band gap of 1.5–2 eV[4–6], making it particularly suitable for practical device integration.

Despite immense research fueled by these favorable attributes, the intrinsic electrical conductivity of CrSBr remains poorly understood. In nominally undoped semiconductors, intrinsic defects are typically the prime source of electronic carriers (electrons and holes). Prior studies have noted that vapor-grown CrSBr can harbor high defect concentrations, with bromine vacancies ($V_{Br}$) believed to be the most abundant species[7] and suspected to act as the origin of *n*-type doping[8]. Additionally, electron transport experiments highlight the interplay between defects and charge carriers in bulk CrSBr[9]. However, a comprehensive and systematic investigation of native defects and their doping behavior has been lacking.

In this work, we combine first-principles density functional theory (DFT) calculations with angle-resolved photoemission spectroscopy (ARPES). In particular, we systematically investigate the thermodynamic and electronic properties of native defects in CrSBr, including vacancies, self-interstitials, and antisites. To qualify as a favorable *n*-type dopant, a defect must have a low formation energy when the Fermi level lies near the conduction band minimum (CBM) and possess stable positive charge states over a significant fraction of the bandgap. In contrast to previous assumptions, our calculations suggest chromium interstitials ($Cr_i$) as energetically favorable defects under *n*-type conditions and capable of acting as shallow donors. Two bromine-related defects, $V_{Br}$ and bromine antisites ($Br_S$), are also identified as potential donors but with larger ionization energies. These findings provide a microscopic explanation of unintentional *n*-type doping in CrSBr and offer guidance for future defect engineering on this promising material.

# Results

## ARPES measurements

Figure 1 presents our ARPES measurements on bulk CrSBr crystals at room temperature (see Methods). In Figure 1a, the ARPES intensity below the valence band maximum (VBM) closely reproduces earlier reports[4,6]. The corresponding energy distribution curve (EDC), shown on the right, reveals a weak but distinct spectral feature above the VBM, highlighted in the inset. We extract an energy separation of $(1.48 \pm 0.09)$ eV between the VBM and the onset of this additional spectral weight by linearly extrapolating the EDC and fitting the unoccupied tail with a Gaussian. Due to low signal strength, the overlaid data above -0.85 eV was acquired with 25 times longer acquisition time.

The observed intensity might stem from mid-gap states created by defects at the surface. J. Klein et al. observed an increased density of $V_{Br}$ at the surface using STM[7], and M. Weile et al. also found a plethora of defect complexes in CrSBr bilayer with $V_{Br}$ being the most abundant[19]. However, DFT calculations revealed rather uniform and flat dispersion of these defect states[19] in stark contrast to the conduction band asymmetry and intensity distribution observed in Figure 1a.

Further insight is obtained from the isoenergetic contour along the in-plane momenta (also known as “momentum map”) shown in Figure 1b, recorded over the energy interval [–1.6, –1.3] eV, corresponding to the region of the additional spectral weight above the VBM. The spectral intensity is highly anisotropic and sharply localized along the $\bar{\Gamma}$-$\bar{X}$ direction. The resulting contour exhibits a conduction band–like topology, consistent with prior observations in ultrathin CrSBr flakes on metallic substrates[10]. Together with the ~1.5 eV bandgap, these features provide direct spectroscopic evidence of partial conduction band occupation.

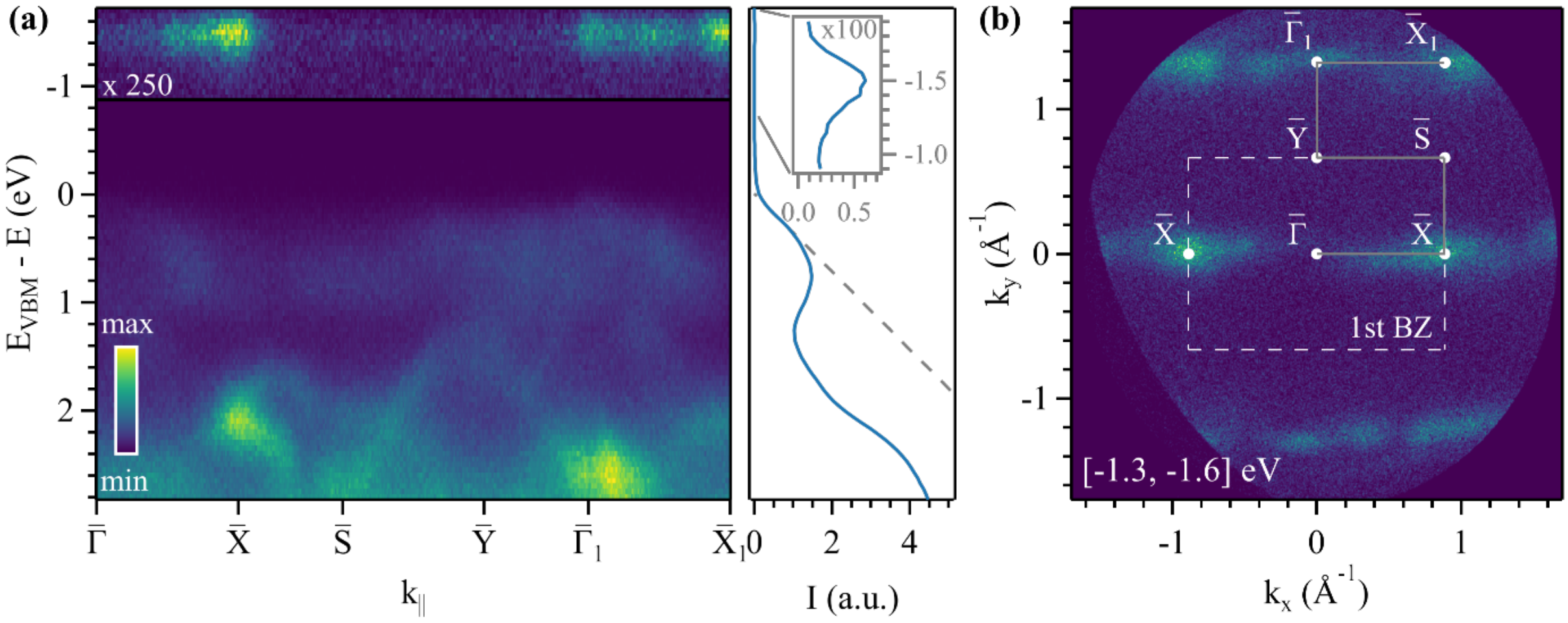

**Figure 1**. **ARPES measurements on bulk CrSBr indicating partial conduction band occupation.** (a) ARPES intensity map of bulk CrSBr at room temperature measured along high symmetry points. Above -0.85 eV, the acquisition time is increased 25-fold and the contrast is increased tenfold. The corresponding energy distribution curve (EDC), shown on the right, displays a weak spectral feature above the valence band maximum (VBM), highlighted in the inset (intensity multiplied by 100 for clarity). The grey dotted line indicates the linear fit used to extrapolate the VBM position. (b) Momentum map integrated over the energy range –1.6 eV to –1.3 eV, corresponding to the energy

window of the conduction band–like feature. The white dotted line marks the boundary of the first Brillouin zone and the grey solid line marks the path displayed in (a).

Crucially, while previous ARPES and transport studies reported similar conduction band signatures, they were attributed to substrate-induced charge transfer in exfoliated flakes[10] or Li doping[11]. In contrast, our measurements were performed on bulk, millimeter-thick crystals, ruling out substrate effects. This strongly supports an intrinsic origin of the conduction band filling, most likely due to native donor-type defects in as-grown CrSBr. This is an important step toward validating the microscopic doping mechanism and establishes a foundation for scalable device integration of CrSBr in substrate-free environments.

## DFT modeling of native point defects in bulk CrSBr

To identify the intrinsic point defects responsible for the observed conduction band population - and thereby the unintentional *n*-type conductivity - we performed first-principles calculations within the framework of DFT. Native defect formation energies depend critically on the atomic chemical potentials of Cr, S, and Br ($\mu_{Cr}$, $\mu_S$, $\mu_{Br}$), which are bounded by the thermodynamic stability region of CrSBr relative to decomposition into competing binary and ternary phases. Therefore, we began our analysis with a thermodynamic assessment of bulk CrSBr. The chemical potentials of the constituting elements are defined as $\mu_i = \mu_i^0 + \Delta\mu_i$, where $\mu_i^0$ is the energy per atom of element *i* in its reference state (Cr in pure bcc chromium crystal, S in $\alpha$-$S_8$ crystal, and Br in orthorhombic bromine crystal), and $\Delta\mu_i \leq 0$. The thermodynamic stability of CrSBr is defined by the equality:

$$\Delta\mu_{Cr} + \Delta\mu_S + \Delta\mu_{Br} = \Delta H_f(\mathrm{CrSBr}), \quad (1)$$

where $\Delta H_f$(CrSBr) is the formation energy of CrSBr per formula unit.

To ensure CrSBr is the most stable phase under given conditions, its formation must be favored over all possible competing phases. Based on our DFT total energy calculations for known entries in the OQMD database[12], we identified $Cr_2S_3$, $Cr_3S_4$, and $CrBr_2$ as the relevant competing phases that constrain the chemical potentials and define the boundaries of the CrSBr stability region. The thermodynamic conditions that prevent decomposition into these phases impose the following inequalities:

$$2\Delta\mu_{Cr} + 3\Delta\mu_S \leq \Delta H_f(\mathrm{Cr_2S_3}), \quad (2)$$

$$3\Delta\mu_{Cr} + 4\Delta\mu_S \leq \Delta H_f(\mathrm{Cr_3S_4}), \quad (3)$$

$$\Delta\mu_{Cr} + 2\Delta\mu_{Br} \leq \Delta H_f(\mathrm{CrBr_2}). \quad (4)$$

We calculate $\Delta H_f$(CrSBr) = –6.09 eV, $\Delta H_f$($Cr_2S_3$) = −9.89 eV, $\Delta H_f$($Cr_3S_4$) = −14.67 eV, and $\Delta H_f$($CrBr_2$) = −6.50 eV per formula unit. All total energy calculations were performed using the PBE+U approach[13,14] (see Methods) that provides reliable formation energies and magnetic ground states, consistent with prior studies of other 2D magnetic semiconductors such as $CrI_3$ and $CrBr_3$[15–17]. Accordingly, from the constraints (1–4), we calculated the stability diagram, which is plotted in Figure 2a in the ($\Delta\mu_S$, $\Delta\mu_{Br}$) space (with $\Delta\mu_{Cr}$ determined from the stoichiometry constraint).

The corners (A, B, C) and the nominal midpoint of the CrSBr stability region marked in Figure 2a define the range of accessible synthesis conditions. These representative points were, therefore, selected for subsequent defect calculations. Point A corresponds to the Cr-rich conditions ($\Delta\mu_{Cr} = -3.1$ eV, $\Delta\mu_{S} = -1.2$ eV, $\Delta\mu_{Br} = -1.6$ eV), point B to the S-rich conditions ($\Delta\mu_{Cr} = -4.9$ eV, $\Delta\mu_{S} = 0.0$ eV, $\Delta\mu_{Br} = -1.1$ eV), and point C to the S-rich and Br-rich conditions ($\Delta\mu_{Cr} = -5.7$ eV, $\Delta\mu_{S} = 0.0$ eV, $\Delta\mu_{Br} = -0.4$ eV).

We investigated eight native defects that we expect to be most relevant: three vacancies ($V_{Cr}$, $V_{S}$, $V_{Br}$), three self-interstitials ($Cr_i$, $S_i$, $Br_i$), and two antisites ($S_{Br}$, $Br_{S}$). Defect calculations were performed in a 4×3×2 supercell (two CrSBr layers) with antiferromagnetic ordering preserved. First, we computed the formation energies, $\Delta H_f[D^0]$, of the defects in their neutral charge states:

$$\Delta H_f\,[D^0] = E_{tot}[\mathrm{CrSBr}{:}D^0] - E_{tot}[\mathrm{CrSBr}] + \sum_i n_i\mu_i, \qquad (5)$$

where $E_{tot}[\mathrm{CrSBr}{:}D^0]$ and $E_{tot}[\mathrm{CrSBr}]$ are the total energies of the defective and pristine supercells, respectively, $n_i$ is the number of atoms of species $i$ added (positive) or removed (negative) to form the defect, and $\mu_i$ is the chemical potential of the corresponding species as defined earlier.

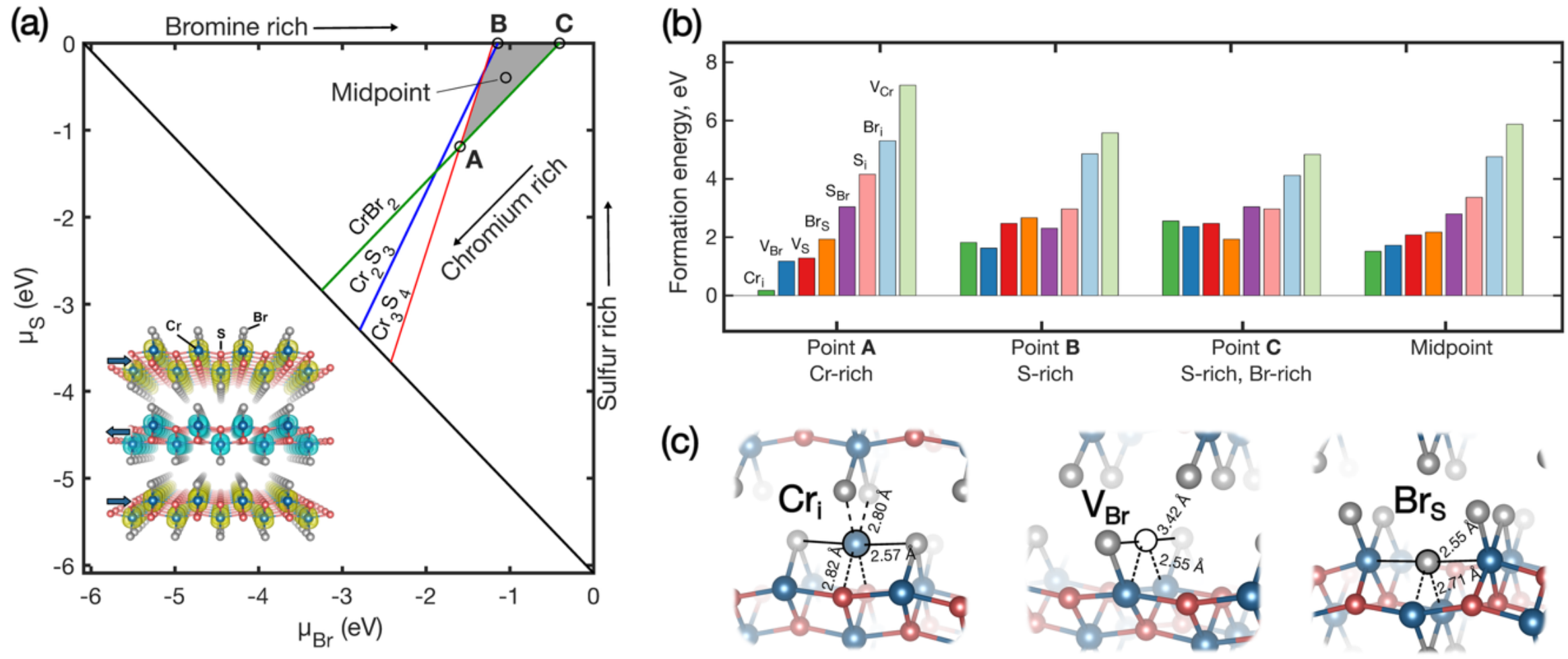

**Figure 2. Formation energies and local atomic structures of neutral intrinsic defects in CrSBr.** (a) DFT-calculated range of chemical potentials (in eV) of the elements in CrSBr. The points A, B, and C mark the representative corners of the stability region of bulk CrSBr (grey area). The inset sketches the atomic structure of CrSBr, with yellow and blue isosurfaces signifying positive and negative magnetization density of the AFM ground state. (b) Formation energies of native point defects in CrSBr in their neutral charge states calculated for the representative points marked in (a). (c) DFT-optimized atomic structures of the most energetically favorable native defects: $Cr_i$, $V_{Br}$, and $Br_S$. For $V_{Br}$, the depicted interatomic distances are measured from the nominal lattice site.

The results are presented in Figure 2b, showing the variation of neutral defect formation energies across the selected chemical environments. The atomic configurations of the most relevant defects are shown in Figure 2c. Among the self-interstitials, $Cr_i$ consistently exhibit favorably low formation energies, particularly under Cr-rich conditions (Point A), where the formation energy falls well below that of other defect types. The relaxed structure of $Cr_i$, shown in Figure 2c, reveals that the most favorable incorporation site is located in the van der Waals gap between the CrSBr layers. The interstitial Cr atom adopts a quasi-octahedral coordination with surrounding atoms, forming bonds of 2.57–2.82 Å to nearby sulfur and bromine atoms. This configuration results in minimal distortion to the adjacent CrSBr layers, preserving the host lattice integrity. Notably, the sulfur and bromine interstitials ($S_i$ and $Br_i$) are also stabilized at interlayer sites, but with much higher formation energies.

Previously, $Cr_i$ has been proposed as a structurally and magnetically significant defect in $He^+$ ion-irradiated CrSBr crystals, where it was found to bridge neighboring layers and stabilize interlayer ferromagnetic coupling[18]. Our calculations show that this defect can favorable as well in as-grown CrSBr under a wide range of growth conditions.

At the same time Figure 2b illustrates that $Cr_i$ does not dominate at all corners of the phase space. Specifically, two vacancies, $V_{Br}$ and $V_S$, show low formation energies in Cr-poor environments. Our results indicate that $V_{Br}$ is expected to be the most favorable defect under S-rich conditions (Point B of the phase diagram), which is consistent with the observations of high concentration of this defect in previous experimental works [7,8]. Its incorporation causes relatively small local perturbation with slight inward relaxations of neighboring atoms (as shown for $V_{Br}$ in Figure 2c).

In contrast to $V_{Br}$ and $V_S$, Cr vacancies ($V_{Cr}$) remain the least favorable of the studied defects, with formation energies consistently around 5–7 eV. At the same time, $V_{Cr}$ remains a noteworthy species since a recent study of defect complexes in CrSBr pointed out its propensity to form Frenkel pairs with $V_{Cr}$.[19] We calculated the formation energy of the most favorable configuration of $V_{Cr}$-$Cr_i$ revealed in Ref. [19] (dubbed $V_{Cr}$-$Cr_i^{\delta 2}$) using the same settings as for other point defects in Figure 2 and found it to be of 3.69 eV – approximately halfway between the individual $V_{Cr}$ and $Cr_i$. It should be noted this does not exclude the appearance of this Frenkel defect pair under non-equilibrium conditions (due to kinetic effects), consistent with microscopy observations[19].

Another noteworthy defect is bromine-on-sulfur antisite, $Br_S$, which becomes the dominant defect in the Br-rich corner of the CrSBr stability region (Point C in Figure 2). Positioned at the sulfur crystallographic site (Figure 2c), the bromine atom of $Br_S$ exhibits modest bond length changes relative to the pristine lattice, with the distance to nearest-neighbor Br atoms increasing from 2.42 Å to 2.55 Å within the same Br sublayer, and from 2.40 Å to 2.72 Å to Br atoms in the adjacent sublayer. These results suggest that bromine atoms introduced, for instance, upon the formation of bromine vacancies are more likely to substitute sulfur atoms within the CrSBr lattice rather than occupy interstitial positions. In contrast to $Br_S$, the second considered antisite defect – sulfur-on-bromine, $S_{Br}$ – exhibits a significantly higher formation energy, although still consistently lower than that of sulfur interstitials ($S_i$) across most of the CrSBr stability region.

The neutral charge state formation energy calculations thus identify two bromine-related defects, $V_{Br}$ and $Br_S$, and the Cr interstitial as the most thermodynamically accessible intrinsic point defects in CrSBr. To further evaluate which defects are electronically active and capable of contributing to *n*-type conductivity, we then examined the charge transition levels (CTLs) of all intrinsic point defects. We used the Slater-Janak (SJ) method[20], where the charge transition levels are estimated from the Kohn–Sham eigenvalues at fractional occupations. The results of the SJ approach are shown to be in good agreement with those obtained from total energy differences, while avoiding the difficulties associated with electrostatic corrections[21,22]. This approach allows for the calculation of CTLs without comparing the total energies of differently charged supercells. While the PBE+U functional is suitable for structural relaxation and estimating formation energies, it significantly underestimates the band gap of CrSBr (PBE+U: 0.62 eV). Therefore, single-point SCF calculations with more accurate exchange-correlation functionals were employed to evaluate CTLs on PBE+U-relaxed geometries. Specifically, we tested the HSE06 hybrid functional[23] and the meta-GGA SCAN functional[24].

It should be noted that the precise value of the band gap in CrSBr remains under debate. Scanning tunneling spectroscopy (STS) and microscopy measurements have reported a gap of approximately 1.5 eV[2,5], whereas earlier ARPES studies suggested a larger values. Reported VBM lie 1.6-1.8 eV[4], 1.5 eV[6], and 1.84 eV[25] below the Fermi level. These earlier investigations, conducted on bulk crystals similar to those used in our study, did not report any spectral signature associated with the conduction band minimum. In contrast, our measurements reveal a faint spectral feature approximately 1.5 eV above the valence band, whose dispersion is consistent with a partially occupied CBM. The spread in reported VBM-Fermi separations may arise from band bending at the surface that shifts the observed valence band position depending on *n*-type doping level, photon energy, and photon fluence[26,27], thus influencing the VBM position and the appearance of a conduction-band signal in the ARPES spectra[26,28].

Beyond the magnitude of the band gap, its character is also contested. Although CrSBr is commonly regarded as a direct-gap semiconductor, two ARPES studies have reported an indirect band gap, one following K dosing[25] and another attributed to substrate-induced charge transfer in exfoliated flakes[10]. In our measurements the marked $\overline{\Gamma}$ point of the first Brillouin zone (BZ) shows weak intensity in Figure 1, whereas neighboring BZs $\overline{\Gamma}_1$ displays strong spectral weight. This effect was previously noted by M. Bianchi et al.[6] and is attributed to photoemission matrix element variations. We therefore compared multiple BZs carefully and found no consistent evidence for an indirect gap with the CBM at the $\overline{X}$ point.

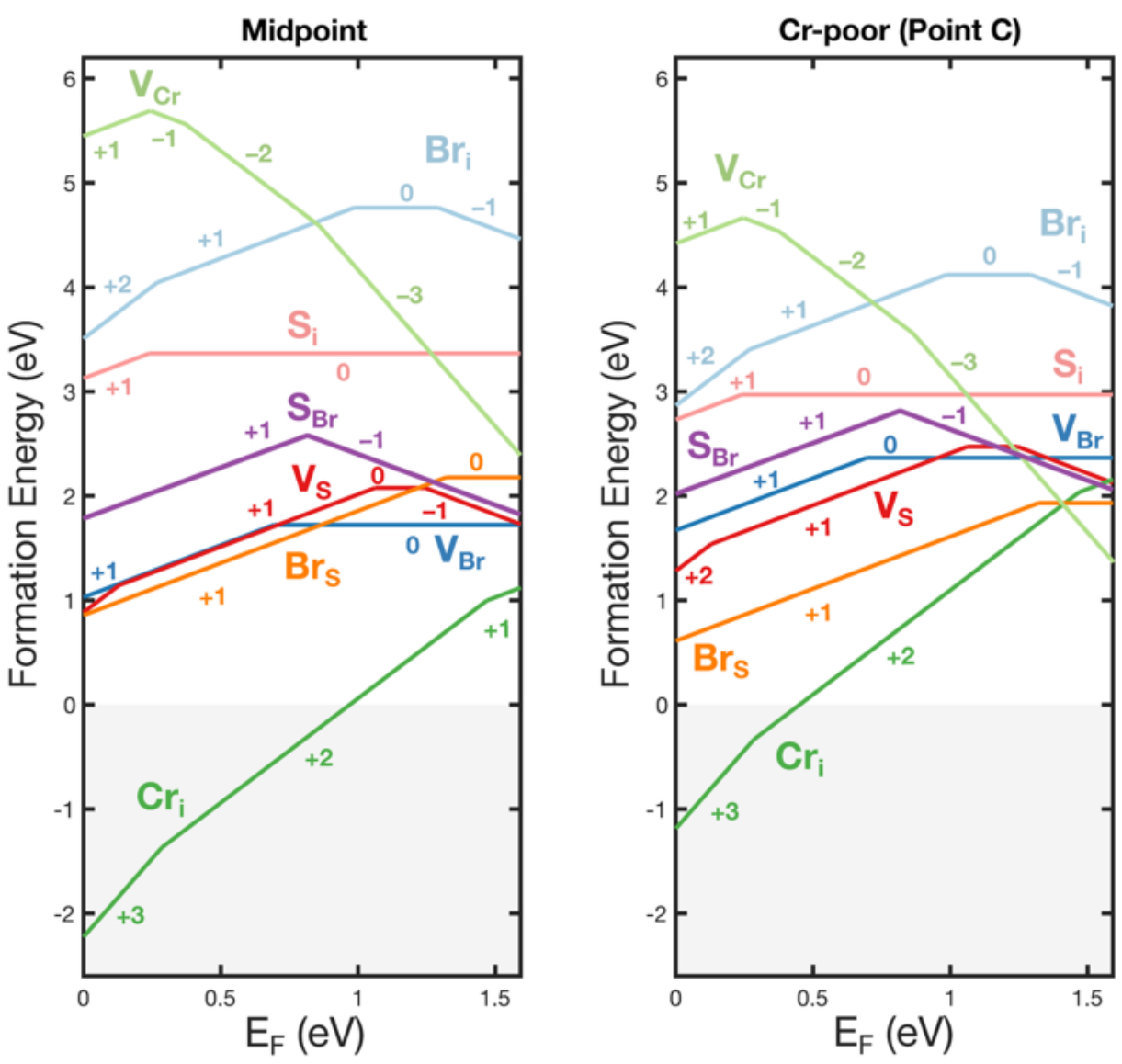

**Figure 3. Defect formation energies as a function of Fermi energy.** Formation energies of intrinsic point defects in CrSBr (as in Figure 1b) in different charge states calculated as a function of Fermi energy for different chemical environments: the midpoint of the CrSBr stability region in Fig. 2a and Cr-poor conditions (Point C in Figure 2a).

In our DFT calculations, the HSE06 hybrid functional yields a band gap of 2.21 eV, while the SCAN meta-GGA functional gives 1.59 eV. Given that the SCAN result closely matches recent transport and STS measurements – as well as the gap inferred from our ARPES data (see also Figure S1) – we adopted SCAN for computing CTLs in our defect analysis.

The calculated CTLs are shown in Figure 3 for the midpoint of the CrSBr stability region and the Cr-poor, S-rich, Br-rich limit (Point C in Figure 2a). The total spin of each energetically favorable charge state is listed in Table S1. Our results suggest that among the studied defects $Cr_i$ emerges as a potent shallow donor. It is stable only in positive charge states (+3, +2, and +1) throughout the entire band gap, with the (+1/0) transition level lying above the CBM, indicating a stable +1 state under *n*-type conditions. Its formation energy at *n*-type conditions is favorable over other studied defects across the majority of the CrSBr stability region (as represented by the midpoint data in Figure 3) save for the Cr-poor conditions (such as Point C).

Other potential donor candidates include $V_{Br}$ and $Br_S$: their formation energy curves indicate positive charge states are favored at low Fermi levels, with transitions to the neutral charge state occurring high in the band gap. The bromine vacancy – a defect whose high concentration in CrSBr is reported in several experimental works[7,8,19] – is stable in +1 and 0 charge states. However, its (+1/0) transition level lies ~0.89 eV below the CBM, implying high ionization energy and classifying it as a deep donor. In contrast to $V_{Br}$, the $Br_S$ antisite has a relatively shallow donor level located ~0.27 eV below the CBM. Moreover, Figure 3 shows that the computed formation energy of $Br_S$ is lower than that of $Cr_i$ in the Cr-poor region under *n*-type conditions (i.e., when the Fermi level is close to the CBM). These observations highlight $Br_S$ and $Cr_i$ as the most likely contributors to unintentional *n*-type doping and the observed conduction-band filling in ARPES, with $V_{Br}$ being the next best candidate.

## XPS measurements and DFT analysis

Finally, we analyzed the consistency of the predicted defects with experimental X-ray photoelectron spectroscopy (XPS) data. For this, we computed the Cr $2p_{3/2}$ core-level binding energies for both defect-containing and bulk supercells, referenced to the energy of the highest occupied state. The excited Cr atoms were described by a pseudopotential generated with a $2p$ core hole. Accordingly, the binding energies were obtained as the total energy differences between the core-hole state and the ground state of the (N − 1)-electron system[29]. To account for the pseudopotential formalism used in our calculations, the computed binding energies were supplemented by an additive correction defined as the difference of all-electron and pseudopotential binding energies of the isolated Cr atom (for details, see Refs. [30, 31]). This yields a binding energy of 577.1 eV for Cr atoms in bulk CrSBr. In the supercell containing $Cr_i$ (in the +1-charge state), the core-level binding energies of bulk atoms remained consistent with this value, while the interstitial Cr atom exhibits a binding energy shifted by about 1 eV, to 576.1 eV.

For comparison, we have also computed the $2p_{3/2}$ core-level binding energy of the $Cr_i$ atom in other relevant charge states. The calculated values are 576.31 eV and 576.32 eV for $Cr_i^{2+}$ and $Cr_i^{3+}$, respectively, comparable with the +1-charge state of $Cr_i$ and shifted by about 0.8 eV from the bulk value. Notably, the $2p_{3/2}$ binding energies of Cr atoms near other energetically favorable defects, $V_{Br}^0$ and $Br_S^0$, are much closer to those of bulk atoms (shifted by 0.2–0.3 eV): for $V_{Br}^0$, the calculated binging energy of the nearest-neighbor chromium is 576.8 eV, and for $Br_S^0$, it is 576.9 eV.

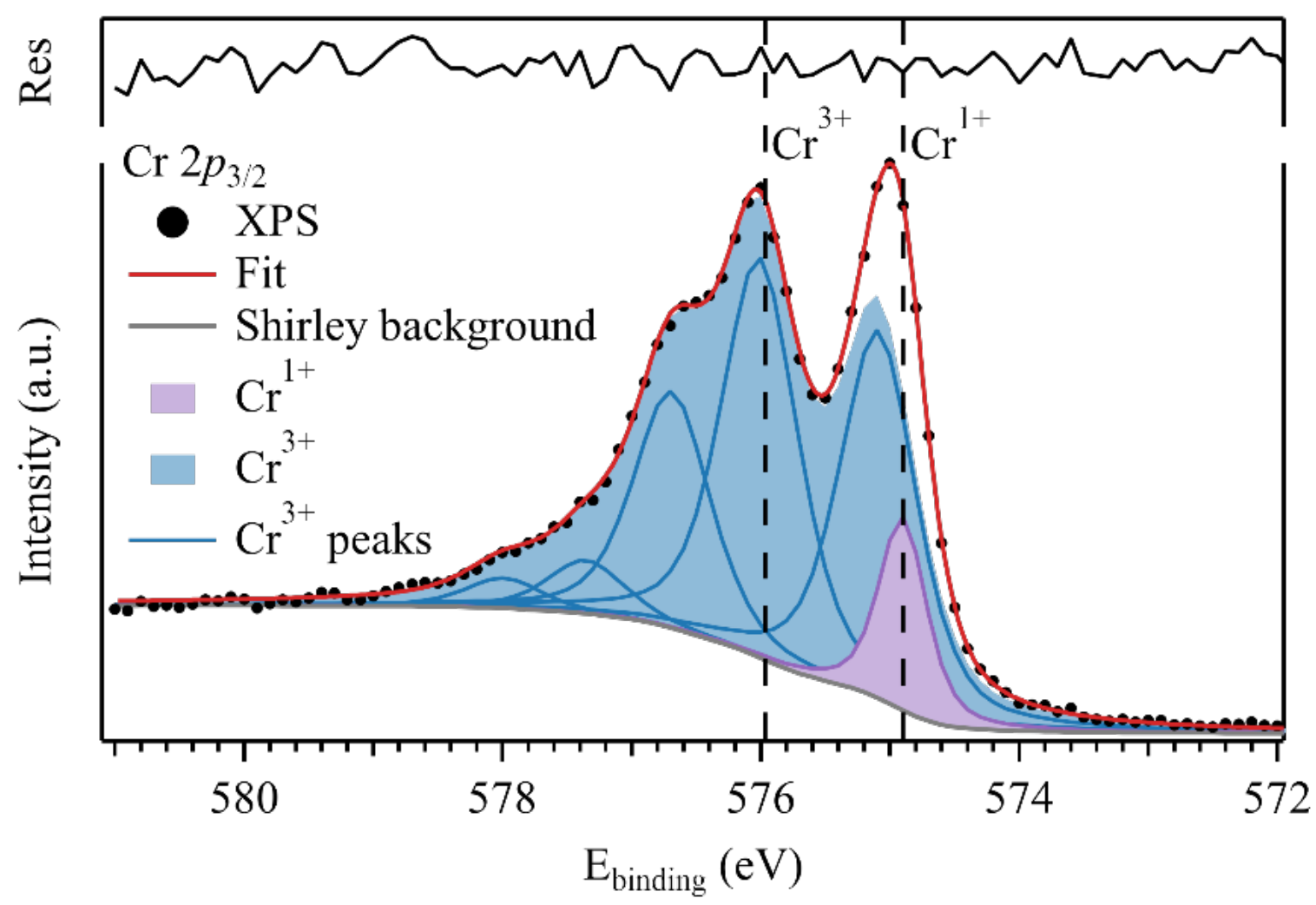

**Figure 4. XPS spectrum of the Cr 2$p_{3/2}$ core level.** The $Cr^{3+}$ contribution from bulk CrSBr is modeled as a multiplet comprising five pseudo-Voigt components (blue), while the $Cr^{1+}$ interstitial is fitted with a single pseudo-Voight peak (purple). A standard Shirley background is included in the overall fit. The combined fit reproduces the experimental data with a flat residual. The main $Cr^{3+}$ peak position is determined via an area-weighted average of the five multiplet components, resulting in a binding energy difference of 1.07(08) eV relative to the $Cr^{1+}$ state (compare dashed lines).

The experimental XSP spectrum of the Cr 2$p_{3/2}$ core level is shown in Figure 4. In addition to the spin-orbit splitting of the 2$p_{1/2}$ and 2$p_{3/2}$ levels, additional multiplet splitting arises from the coupling between the remaining 2$p_{3/2}$ core electron and the 3*d* valence electrons[32]. Even though a multiplet structure is evident by eye in the experimental data, previous studies on CrSBr have neglected this effect[32,33], potentially leading to misassignment of the Cr binding energy[34,35]. In our analysis, we fit the $Cr^{3+}$ component using a multiplet of five pseudo-Voight functions and include an additional pseudo-Voight peak to account for the $Cr^{1+}$ interstitials. We applied a standard Shirley background and all resulting fit parameters are summarized in Table S2. The fit reveals a composition of 90.1 % $Cr^{3+}$ and 9.9 % $Cr^{1+}$. We determine the $Cr^{3+}$ binding energy to be 575.97(04) eV by calculating an area-weighted average of the five multiplet components, while the $Cr^{1+}$ peak is centered at 574.90(04) eV, yielding a splitting of 1.07(08) eV. The Br 3*d* and S 2*p* core level spectra are shown in Figure S2 and support a singular charge state for both elements. Although the absolute values of the Cr peaks differ slightly from our DFT results, the agreement in the relative binding energy shift between the two oxidation states aligns with the presence of interstitial Cr atoms in the +1-charge state. While XPS cannot by itself unambiguously identify the defect species, this close correspondence between theory and experiment suggests that the $Cr_i$ defects predicted by our DFT analysis are indeed realized in as-grown CrSBr.

## Discussion

Our combined experimental and DFT results underscore native point defects as possible sources of intrinsic electron doping in CrSBr. The chromium self-interstitial, $Cr_i$, bromine-on-sulfur antisites, $Br_S$, and bromine vacancies, $V_{Br}$, are identified as the most likely intrinsic shallow donor. Importantly, our defect formation energy analysis provides guidance for growth optimization: Cr-poor, Br-rich, and S-rich conditions are expected to suppress the formation of these donor defects and promote intrinsic behavior. This identification, and the comprehensive analysis of defect chemistry in CrSBr presented here, opens new avenues for controlling defect formation and achieving truly intrinsic or intentionally doped states. Establishing control over native doping is not only vital for fundamental studies of CrSBr's coupled electronic and magnetic phenomena, but also for optimizing this 2D magnet in future applications. In addition, our ARPES measurements provide direct spectroscopic evidence that CrSBr is a direct band gap semiconductor, further confirming its suitability for optoelectronic and spintronic applications.

# Methods

***Sample Preparation***. We used commercial CrSBr crystals from HQ Graphene, which we glued to copper sample plates using UHV-compatible silver epoxy (EPO-TEK H21D) to ensure electrical grounding. After transfer into the UHV system, the crystals were exfoliated in situ using Scotch tape to obtain a clean surface. Exfoliation was carried out at a base pressure better than $3\times10^{-7}$ mbar for ARPES measurements and $> 5\times10^{-8}$ mbar for XPS to expose a clean surface.

***ARPES***. We performed μ-ARPES using a Kreios 150 MM momentum microscope (Specs GmbH) under UHV conditions ($< 3\times10^{-10}$ mbar)[36]. This microscope allows to acquire two-dimensional isoenergetic maps at a constant kinetic energy, with a momentum field of view spanning approximately $k_x$, $k_y \in [-2.0, +2.0]$ $Å^{-1}$. Photoelectrons were excited using the He Iα resonance line (21.2 eV) from a Helium discharge lamp equipped with a monochromator, providing a spot size of approximately 200 μm in diameter. This configuration allows probing of locally flat regions, suitable for μ-ARPES measurements. All spectra were acquired at room temperature with an energy resolution of 90 meV.

***XPS***. XPS spectra were acquired at room temperature with a Phoibos 150 hemispherical analyzer (Specs GmbH) under UHV conditions. A monochromatized Al Kα X-ray source with 1486.8 eV photon energy facilitated photoemission at 430 meV energy resolution. The binding energy scale was calibrated using the Fermi edge of a clean Au(111) crystal. Spectral analysis was conducted using the XPST package for Igor Pro[37], employing a Gauss-Lorentzian sum function to approximate a Voigt profile and a standard Shirley background for inelastic scattering correction.

***DFT Calculations***. All calculations were performed within the spin-polarized DFT formalism using the Quantum ESPRESSO program[38,39]. Geometry optimization and total energy calculations were conducted with the PBE+U functional[13,14], where the Coulomb energy $U = 5.1$ eV defined self-consistently[40] was applied to Cr 3*d* orbitals to correct for strong on-site correlations. To obtain the band structure and CTLs, we employed the meta-GGA SCAN functional[21] (see the main text). We used norm-conserving pseudopotentials with the kinetic energy cutoff of 1100 eV imposed to achieve converged results.

# Acknowledgements

The Momentum Microscope has been financed by the Deutsche Forschungsgemeinschaft (DFG) through the project INST 212/409 and by the "Ministerium für Kultur und Wissenschaft des Landes Nordrhein-Westfalen". We acknowledge Paderborn Center for Parallel Computing (PC2) for the provided computational resources. We acknowledge financial support by the DFG through project 231447078 (TRR 142/3 – projects A08 and B07) and from the European

Union's Horizon 2020 Research and Innovation Programme under Project SINFONIA, grant 964396.

## Conflict of Interest

The authors have no conflicts to disclose.

## Declaration of Generative AI and AI-assisted Technologies

During the preparation of this work, the authors used ChatGPT to enhance the quality of writing by improving grammar, style, and clarity. After using this tool/service, the authors reviewed and edited the content as needed and take full responsibility for the content of the publication.

## Author Contributions

T.B., W.G.S., K.J.S., A.I., and M.Ci. conceptualized and planned the project and contributed to the writing of the manuscript. T.B. performed DFT calculations. K.J.S., J.E.N., L.S., and G.Z. conducted ARPES measurements. K.J.S., M.S.A., and M.Ca. conducted XPS measurements. W.G.S. and M.Ci. supervised the project.

## Data Availability

The data supporting the findings of this study are available within the article. Additional data related to this manuscript may be requested from the authors.

**Figure 1**. **ARPES measurements on bulk CrSBr indicating partial conduction band occupation.** (a) ARPES intensity map of bulk CrSBr at room temperature measured along high symmetry points. Above -0.85 eV, the acquisition time is increased 25-fold and the contrast is increased tenfold. The corresponding energy distribution curve (EDC), shown on the right, displays a weak spectral feature above the valence band maximum (VBM), highlighted in the inset (intensity multiplied by 100 for clarity). The grey dotted line indicates the linear fit used to extrapolate the VBM position. (b) Momentum map integrated over the energy range –1.6 eV to –1.3 eV, corresponding to the energy window of the conduction band–like feature. The white dotted line marks the boundary of the first Brillouin zone and the grey solid line marks the path displayed in (a).

**Figure 2. Formation energies and local atomic structures of neutral intrinsic defects in CrSBr.** (a) DFT-calculated range of chemical potentials (in eV) of the elements in CrSBr. The points A, B, and C mark the representative corners of the stability region of bulk CrSBr (grey area). The inset sketches the atomic structure of CrSBr, with yellow and blue isosurfaces signifying positive and negative magnetization density of the AFM ground state. (b) Formation energies of native point defects in CrSBr in their neutral charge states calculated for the representative points marked in (a). (c) DFT-optimized atomic structures of the most energetically favorable native defects: $Cr_i$, $V_{Br}$, and $Br_S$. For $V_{Br}$, the depicted interatomic distances are measured from the nominal lattice site.

**Figure 3. Defect formation energies as a function of Fermi energy.** Formation energies of intrinsic point defects in CrSBr (as in Figure 1b) in different charge states calculated as a function of Fermi energy for different chemical environments: the midpoint of the CrSBr stability region in Fig. 2a and Cr-poor conditions (Point C in Figure 2a).

**Figure 4. XPS spectrum of the Cr 2*p*$_{3/2}$ core level.** The $Cr^{3+}$ contribution from bulk CrSBr is modeled as a multiplet comprising five pseudo-Voigt components (blue), while the $Cr^{1+}$ interstitial is fitted with a single pseudo-Voight peak (purple). A standard Shirley background is included in the overall fit. The combined fit reproduces the experimental data with a flat residual. The main $Cr^{3+}$ peak position is determined via an area-weighted average of the five multiplet components, resulting in a binding energy difference of 1.07(08) eV relative to the $Cr^{1+}$ state (compare dashed lines).